\documentclass{aa}
\usepackage{graphicx}
\begin{document}

   \title{Multi-level 3D non-LTE computations of lithium lines 
in the metal-poor halo stars HD140283 and HD84937}

\author{Martin Asplund\inst{1}
        \and
        Mats Carlsson\inst{2}
        \and
        Andreas V. Botnen\inst{2}
        }

\titlerunning{3D non-LTE Li line formation in halo stars}
\authorrunning{Asplund, Carlsson \& Botnen}

\offprints{martin@mso.anu.edu.au}

\institute{Research School of Astronomy, 
           Mt. Stromlo Observatory, 
           Cotter Road, Weston, ACT 2611, Australia
           \and
           Institute of Theoretical Astrophysics, University of Oslo,
           P.O. Box 1029, Blindern, N-0315 Oslo, Norway
           }

\date{}

\abstract{
The lithium abundances in metal-poor halo stars are of 
importance for cosmology, galaxy evolution and stellar structure. 
In an attempt to study possible systematic errors in the 
derived Li abundances, the line formation of Li\,{\sc i} lines
has been investigated by means of 
realistic 3D hydrodynamical model atmospheres of halo
stars and 3D non-LTE radiative transfer calculations.
These are the first detailed 3D non-LTE computations reported employing 
a multi-level atomic model showing that such problems are now
computationally tractable.
The detailed computations reveal that the Li\,{\sc i} population
has a strong influence from the radiation field rather than the local
gas temperature, indicating that the low derived Li 
abundances found by Asplund et al. (1999) 
are an artifact of their assumption of LTE.
Relative to 3D LTE, the detailed calculations 
show pronounced over-ionization.
In terms of abundances the 3D non-LTE values are within 0.05\,dex
of the 1D non-LTE results for the particular cases
of HD\,140283 and HD\,84937, which 
is a consequence of the dominance of the radiation in determining
the population density of Li\,{\sc i}. Although 3D non-LTE can be
expected to give results rather close ($\approx \pm 0.1$\,dex) 
to 1D non-LTE for this reason,
there may be systematic trends with metallicity
and effective temperature. 
   
          }

 \maketitle
%

\section{Introduction}

Lithium abundances of metal-poor halo stars have had a prominent place
in astrophysics and cosmology ever since the first discovery of Li 
in such stars 
(Spite \& Spite 1982) due to their wide-ranging implications. 
Firstly, the uniformity of the Li abundances in halo stars
(Ryan et al. 1999) 
presumably reflects the primordial Li abundance stemming from 
Big Bang nucleosynthesis, thus allowing an estimate of the
baryon density of the Universe.
Secondly, any observed slope in the Spite-plateau with metallicity,
when combined with the corresponding behaviour of $^6$Li, Be, B and O,
can be interpreted in terms of Galactic chemical evolution, in particular
the history of cosmic ray spallation in the early Galaxy. Thirdly,
since it is a fragile element,
Li can
function as a tracer of stellar mixing; the thinness of the Spite-plateau
certainly severely limits the allowed amount of Li-depletion in halo stars
(Ryan et al. 1999).

In order to extract Li abundances from an observed spectrum it is
necessary to have
suitable models of the stellar photosphere and the line
formation process. Since traditional abundance analyses utilize
1D hydrostatic model atmospheres based on
local thermodynamic equilibrium (LTE) and
a rudimentary description
of convection in the form of the mixing length theory (MLT), there
may be significant systematic errors, which could distort the derived
conclusions.
The detailed line formation, including departures from LTE but in 1D
hydrostatic model atmospheres, were investigated by Carlsson et al. (1994)
who found relatively small non-LTE abundance corrections ($\la 0.05$\,dex)
for low metallicity stars.
To investigate the effects of convection on the line formation,
Asplund et al. (1999) applied the new generation of 3D, hydrodynamical
model atmospheres to abundance analyses of metal-poor stars 
assuming LTE.
Due to the much lower temperatures encountered in the line-forming layers 
than in classical 1D model atmospheres, 
the derived abundances differ drastically.
Most notably, it was found that the primordial Li abundance may previously
have been over-estimated by $\simeq 0.3$\,dex.
Although such 3D model atmospheres should be a more realistic
description of the stellar photospheres (Asplund et al. 2000), 
the simplification of LTE in the line formation is a 
crucial assumption. Asplund et al. (1999) indeed 
warned that the steep temperature gradients may be prone
to significant over-ionization of species like Li\,{\sc i}. 
The aim of the present paper is to investigate possible departures from
LTE by performing realistic 3D non-LTE calculations for Li\,{\sc i} lines,
while we defer a detailed investigation of the line formation 
to a subsequent article.
Preliminary calculations have been presented in Asplund (2000)
using a smaller Li atom and without consideration of line-blanketing.

\section{3D non-LTE radiative transfer of Li\,I lines}

From the {\it ab-initio}
3D radiative-hydrodynamical convection simulations
of the metal-poor halo stars HD\,140283 and HD\,84937 presented
in Asplund et al. (1999), two representative and independent
snapshots have been selected
from each simulation for the non-LTE calculations.
Additionally, two snapshots were taken from a similar
solar simulation (Asplund et al. 2000a) for comparison purposes.
The original simulation data cubes with dimensions 100$^2$\,x\,82,
were interpolated prior to the non-LTE calculations 
to a 25$^2$\,x\,100 grid with an improved vertical resolution. 
Test calculations with 50$^2$\,x\,100 snapshots 
verified that the procedure had insignificant effect on the
3D non-LTE abundance corrections.
Typically the atmospheric structures cover the region 
$-6 \le {\rm log} \tau_{500} \le 2$.
Further details of the numerical simulations are available
in Stein \& Nordlund (1998) and Asplund et al. (1999, 2000).

The Li model atom 
employed in the present study is identical to the one
compiled by Carlsson et al. (1994).
The adopted 21-level atom
consists of in total 70 
bound-bound and 20 bound-free radiative transitions. 
We feel confident that the employed Li atom is sufficiently extended
as tests with a more restricted 12-level atom revealed non-LTE
abundance corrections only about 0.03\,dex larger than with the
21-level atom for HD\,140283 and HD\,84937.
The input data for level energies, 
radiative and electron collisional transitions
are in all cases the same as in Carlsson et al. (1994);
accordingly, collisions with neutral hydrogen have not been included.
We have verified that even inclusion of H collisions according
to the classical Drawin (1968) recipe has a very minor impact on our results:
for HD\,140283 the abundance corrections are decreased by 0.05\,dex 
and 0.01\,dex for multiplication factors of 1.0 and 0.01, respectively, of
the standard Drawin estimates. 
Additional atomic data such as background opacities 
(including line-blanketing for photo-ionization transitions)
are taken from the {\sc marcs} package (Gustafsson et al. 1975 and
subsequent updates).

\begin{figure}[t!]
\resizebox{\hsize}{!}{\includegraphics{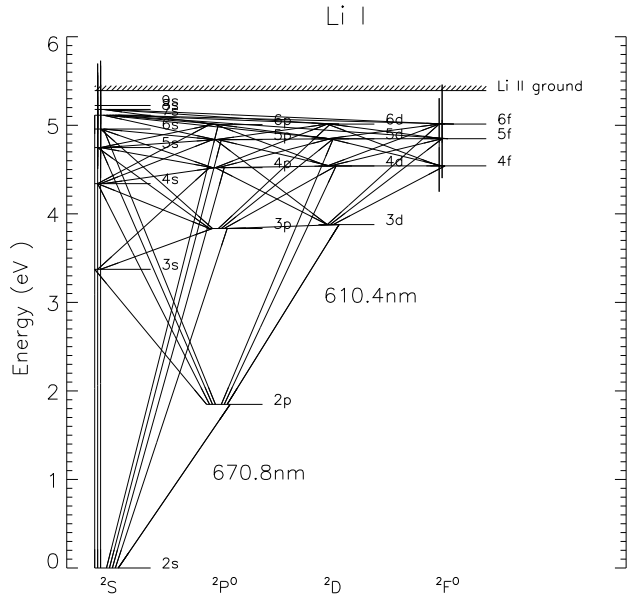}}
\caption{Grotrian termdiagram for the adopted 21-level Li model atom.
All levels are connected with the Li\,{\sc ii} ground state by
photo-ionization transitions. 
}
\label{f:termdiag}
\end{figure}

\begin{table}[t!]
\caption{Predicted LTE and non-LTE flux 
equivalent widths (in pm=10\,m\AA ) of the Li\,{\sc i} 670.8\,nm
line for the Sun, HD\,140283 and HD\,84937 in 1D and 3D. 
The 3D equivalent widths
are the average of two independent snapshots.
\label{t:eqwidth}
}
\begin{tabular}{lccccc} 
 \hline \\
Star & log\,$\epsilon_{\rm Li}$ & 
\multicolumn{2}{c}{1D} & \multicolumn{2}{c}{3D} \\
     &                          &
$W_\lambda^{\rm LTE}$  & $W_\lambda^{\rm NLTE}$ &
$W_\lambda^{\rm LTE}$  & $W_\lambda^{\rm NLTE}$   \\
\hline \\
Sun 		& 1.1 & 0.40 & 0.34 & 0.55 & 0.37 \\
HD\,140283	& 1.8 & 2.38 & 2.18 & 3.84 & 1.96 \\
		& 2.2 & 4.90 & 4.76 & 7.67 & 4.21 \\
HD\,84937	& 2.0 & 1.31 & 1.44 & 1.79 & 1.11 \\
		& 2.4 & 2.91 & 3.29 & 4.00 & 2.55 \\
\hline \\
\end{tabular}
\end{table}

The non-LTE calculations for Li have been performed with 
{\sc multi3d} (Botnen 1997; Botnen \& Carlsson 1999), 
which is essentially a 3D-version of the
widely used {\sc multi}-code for 1D statistical equilibrium problems 
(Carlsson 1986). {\sc multi3d} iteratively
solves the rate equations with a consistent radiation 
field obtained from a simultaneous solution of  
the radiative transfer equation along 24-48 inclined rays.
The formal solution of the radiative
transfer equation is computed using a short characteristic technique,
making use of the horizontal periodic boundary conditions of the hydrodynamical
model atmospheres. The Doppler shifts introduced by the convective motions
are taken into account and therefore no micro- and macroturbulence enter the
analysis (Asplund et al. 2000). 
Typically $\la 6$ accelerated lambda-iterations with linearization and
preconditioning of the rate equations (Scharmer \& Carlsson 1985) 
were necessary to achieve the stipulated convergence criteria 
max$(\delta n_{\rm i}/n_{\rm i}) < 10^{-4}$.
The resulting angle-dependent radiation field
was subsequently used to compute flux profiles. 
Various test calculations with 3D-extended plane-parallel homogeneous
model atmospheres ensured that the same results were obtained with
{\sc multi3d} and {\sc multi}.

\section{Departures from LTE in Li\,I line formation
\label{s:3DNLTE}}

Table \ref{t:eqwidth} summarizes the resulting 3D LTE and non-LTE 
line strengths for the different snapshots and input Li abundances
together with the corresponding 1D results.
Clearly, there are very significant departures from LTE, in particular 
for the two metal-poor stars. 
As predicted by Asplund et al. (1999), over-ionization plays the
dominant role.

\begin{figure}[t!]
\resizebox{\hsize}{!}{\includegraphics{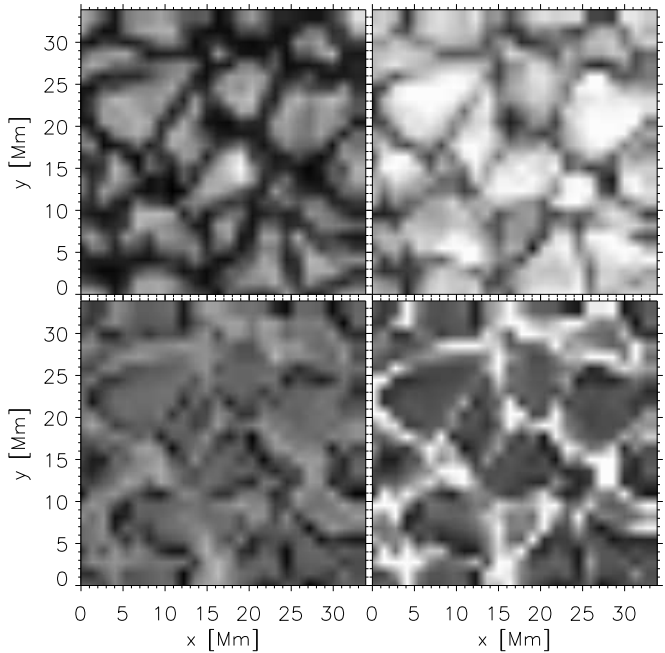}}
\caption{Granulation pattern in snapshot 1 of HD\,140283
seen in disk-center continuum intensity ({\it upper left panel}) and
equivalent width of the Li\,{\sc i} 670.8\,nm line in
LTE ({\it upper right panel}) and non-LTE ({\it lower left panel});
the equivalent width images have the same relative intensity scale to
emphasize the overall difference in line strengths.
Also shown is the ratio of the non-LTE and LTE equivalent widths
({\it lower right panel}).
}
\label{f:IandW}
\end{figure}

Fig. \ref{f:IandW} shows images of the continuum intensity
and the equivalent width of the Li\,{\sc i} 670.8\,nm line
in LTE and non-LTE at disk-center ($\mu = 1.0$).
The characteristic granulation pattern with warm (bright) upflows and cool
(dark) downflows is clearly visible in the continuum intensities.
In LTE the Li line typically is strong above the granules, a consequence of the
low temperatures in the optically thin layers due to the temperature
contrast reversal in the convectively stable layers, 
which translates to a high population density
of Li\,{\sc i}.
Over the downflows the temperatures tend to be higher than average, thus
making the Li line weaker. In non-LTE quite the opposite
happens: in general the line is weaker above the upflows, which implies that the 
population density is
largely controlled by the radiation field rather than the local gas
temperature. 

\begin{figure}[t!]
\resizebox{\hsize}{!}{\includegraphics{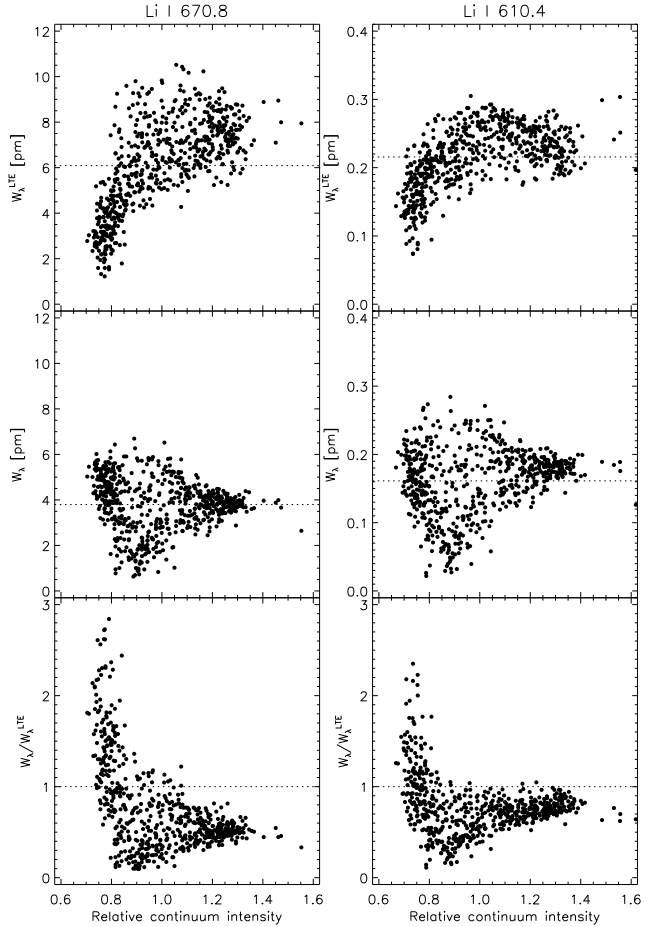}}
\caption{Predicted intensity ($\mu = 1.0$) 
equivalent widths of the Li\,{\sc i} 670.8\,nm {\em (left panel)}
and 610.4\,nm {\em (right panel)} lines across the granulation
pattern in LTE {\em (upper panel)} and non-LTE {\em (middle panel)} 
for the HD\,140283 simulation 
with log\,$\epsilon_{\rm Li} = 2.20$.
The mean intensity equivalent widths are denoted by horizontal lines.
Also shown are the ratios of non-LTE and LTE equivalent widths
{\em (lower panel)}.
}
\label{f:WvsI}
\end{figure}

The same conclusion is obvious from Fig. \ref{f:WvsI},
which quantifies the behaviour in Fig. \ref{f:IandW}. The Li\,{\sc i} 
lines tend to increase in strength with continuum intensity in LTE
but with a very large scatter,
since the temperatures in the higher atmospheric layers
are not one-to-one correlated with the temperatures in the 
continuum-forming layers below. In non-LTE the scatter
is much smaller for regions with high continuum intensities,
which again emphasizes that the line formation process is largely governed by
the non-local properties of the radiation field. 
The 670.8\,nm line is, however, weakest in relatively low continuum intensity
regions ($I^{\rm cont}/\langle I^{\rm cont} \rangle \simeq 0.9$ 
in Fig. \ref{f:WvsI}) 
which are immediately adjacent to very bright regions:
the non-vertical hot radiation field from neighboring granules cause
significant over-ionization (low line opacity) which when coupled
to the shallow temperature gradient of downflows makes the line
very weak (e.g. at $x=21$\,Mm and $y=18$\,Mm in Fig. \ref{f:IandW}). 
The Li\,{\sc i} line typically is strongest in the middle of larger areas
of downflowing material, which has weaker photo-ionizing
radiation field (e.g. at $x=9$\,Mm and $y=13$\,Mm).
The non-LTE behaviour of the Li\,{\sc i} 670.8\,nm line shown
in Figs. \ref{f:IandW} and \ref{f:WvsI}
is similar to the findings of Kiselman (1997, 1998) and 
Uitenbroek (1998) in the case of the Sun, which are confirmed by 
spatially resolved solar observations (Kiselman \& Asplund 2001).
The Li\,{\sc i} 610.4\,nm line show a similar behaviour as 
the resonance line although less dramatically
due to the weakness of the line and its higher excitation potential
(Fig. \ref{f:WvsI}). 

\begin{figure}[t!]
\resizebox{\hsize}{!}{\includegraphics{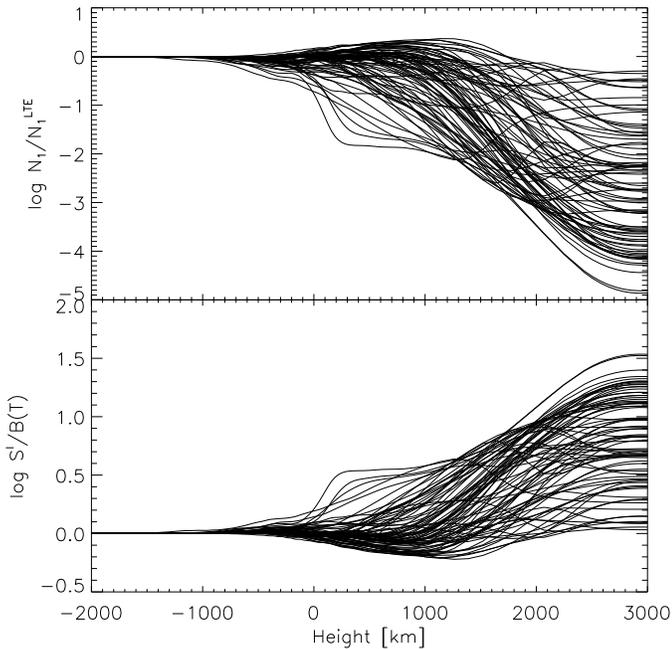}}
\caption{
Variation of the departure coefficient of 
the Li\,{\sc i} ground level $N_1/N_1^{\rm LTE}$ 
{\it (upper panel)} and $S_\nu^{670.8}/B_\nu (T)$ {\it (lower panel)} 
as a function of atmospheric height 
for the HD\,140283 simulation.
Selected vertical columns are connected with solid lines.
}
\label{f:depart}
\end{figure}

Further support for the departures from LTE being essentially an
over-ionization effect comes from the variation with height 
of the departure coefficient of the Li\,{\sc i} ground level 
(Fig. \ref{f:depart}). 
From the LTE-expectation in the deeper layers, the population density 
falls rapidly with height 
as Li\,{\sc i} is photo-ionized
by radiation from below. The non-LTE line formation 
of the Li\,{\sc i} 670.8\,nm line is
mainly an optical depth effect since the line source function $S_\nu^{670.8}$
in general remains reasonably close to the Planck function $B_\nu (T)$
(Fig. \ref{f:depart}).
As in the 1D case, the over-ionization is predominantly driven 
by photo-ionization from the 2p state for which 
$J_\nu/B_\nu$ is near its maximum (Carlsson et al. 1994).
The inclusion of line opacities to the background opacities
for bound-free transitions is thus of importance: without line-blanketing
the non-LTE abundance corrections for HD\,140283 would be about 0.07\,dex
higher. 
As expected the radiative rates dominate over the corresponding 
collisional rates for the 2s and 2p levels.

The speculation by
Kurucz (1995) that the strength of the Li\,{\sc i} 670.8\,nm line
could be seriously overestimated with 1D models due to  
substantial departures from LTE  
in the presence of temperature inhomogeneities, is thus partly
validated by our results, although for the wrong reasons and to a much
lesser extent. Due to his simple two-stream 
MLT-convection approach, Kurucz also failed to recognize 
the compensating effect 
due to convection (Asplund et al. 1999; Asplund 2000).
Cayrel \& Steffen (2000) reached 
similar conclusions to ours based on calculations using a 5-level Li atom
and a 2D simulation of a metal-poor Sun. 

\section{Discussion
\label{s:discussion}}

With the aid of Table \ref{t:eqwidth}, it is possible to interpolate
the computed 1D and 3D line strengths to the observed equivalent widths
of Li\,{\sc i} 670.8\,nm for HD\,140283 (4.65\,pm) and HD\,84937 (2.44\,pm). 
It is noteworthy that the 3D non-LTE abundances only differ by
$\la 0.05$\,dex 
from the corresponding 1D non-LTE
results for HD\,140283 and HD\,84937. 
The 3D LTE abundances on the other hand are distinctly different 
by $0.2-0.3$\,dex.
As cautioned by Asplund et al. (1999), LTE is clearly a very 
poor assumption for Li\,{\sc i} in 3D models of halo stars with their low
surface temperatures and steep temperature gradients
\footnote{This statement is of course not true in general,
as LTE may be a very good approximation for other
species.}. 

The 3D non-LTE abundances are very similar to the 1D non-LTE
results. The main reason for this is that the line strength is
determined by the opacity which is set by the photoionizing radiation
field while collisions are found to be relatively unimportant
according to our calculations\footnote{
We note that the approximate ionization equation (e.g. 
Mihalas 1978, Eq. 5.46) is not applicable as a possible explanation
for the similarity of the 1D and 3D results, since the downward rate
from Li\,{\sc ii}
is dominated by collisions to upper levels 
rather than by photo-recombination to the ground level.}. 
This radiation is optically thin at the height of formation of
the Li\,{\sc i} resonance line and is thus equal to the observable
emergent radiation field.  Since 1D and 3D models
produce similar ($\pm 10$\%) emergent intensities at the
relevant wavelengths (Asplund \& Garc\'{\i}a P{\'e}rez 2001),
the photo-ionization radiation fields will not differ
significantly. The final line strengths will however also 
depend on the details of the temperature structures in the
deep atmospheric layers which differs between 1D and 3D models. 

From the results obtained here for two typical halo stars
we expect that the 3D non-LTE abundances will
be similar ($\approx \pm 0.1$\,dex) 
to the 1D non-LTE results (and to 1D LTE given the small
1D non-LTE abundance corrections according to Carlsson et al. 1994) 
in general for halo stars. 
However, given the great interest in accurately determining
the primordial Li abundance, the intrinsic scatter in the
Spite-plateau and the possible existence of a trend with
metallicity for the Li abundances (e.g. Ryan et al. 1999),
it is clearly important to extend the 3D non-LTE calculations
to additional stars with different parameters
to determine exactly the net 3D effects. 
We are currently working towards this goal 
(Asplund et al., in preparation).

\acknowledgements
It is a pleasure to thank R. Cayrel, R. Collet, 
A.E. Garc\'{\i}a P{\'e}rez and D. Kiselman for stimulating discussions
and the referees for helpful comments.
This work has been supported by grants from the
Swedish and Norwegian Research Councils.


\begin{thebibliography}{}
\bibitem[]{} Asplund, M. 2000, in IAU Symp. 198,
The light elements and their evolution, ed. L. Da Silva et al.,
448 
\bibitem[]{} Asplund, M., \& Garc\'{\i}a P{\'e}rez, A.E. 2001, A\&A, 372, 601
\bibitem[]{}  Asplund, M., Nordlund, \AA., Trampedach, R., \& Stein, R.F., 
1999, A\&A, 346, L17
\bibitem[]{}  Asplund, M., Nordlund, \AA., Trampedach, R., Allende Prieto, C., \&
Stein, R.F. 2000, A\&A, 359, 729
\bibitem[]{} Botnen, A.V. 1997, Cand. Sci. Thesis, University of Oslo
\bibitem[]{} Botnen, A.V., \& Carlsson, M. 1999, in 
Numerical astrophysics, ed. S.M. Miyama et al.,
379
\bibitem[]{} Carlsson, M. 1986, Uppsala Astronomical Report No. 33
\bibitem[]{} Carlsson, M., Rutten, R.J., Bruls, J.H.M.J., \& Shchukina, N.G.
1994, A\&A, 288, 860 
\bibitem[]{} Cayrel, R., Steffen, M. 2000, in IAU Symp. 198, 
The light elements and their evolution, ed. L. Da Silva et al.,
437 
\bibitem[]{} Drawin, H.W. 1968, Z. Phys., 211, 404
\bibitem[]{} Gustafsson, B., Bell, R.A., Eriksson, K., \& 
Nordlund, \AA . 1975, ApJ, 42, 407
\bibitem[]{} Kiselman, D. 1998, A\&A, 333, 732
\bibitem[]{} Kiselman, D. 1997, ApJ, 489, L107
\bibitem[]{} Kiselman, D., \& Asplund M. 2001, in ASP Conf. Ser. 223, 
Cool stars, stellar systems and the Sun, 
ed. R.J. Garc\'{\i}a L\'opez et al., 684 
\bibitem[]{} Kurucz, R.L. 1995, ApJ, 452, 102
\bibitem[]{} Mihalas D., 1978, Stellar atmospheres, W.H. Freeman and 
Company, San Fransisco
\bibitem[]{} Ryan, S.G., Norris, J.E., \& Beers, T.C. 1999, ApJ, 523, 65
\bibitem[]{} Scharmer, G.B., \& Carlsson, M. 1985, J. Comp. Phys., 59, 56 
\bibitem[]{} Spite, F., \& Spite, M. 1982, A\&A, 115, 357
\bibitem[]{} Stein, R.F., \& Nordlund, \AA. 1998, ApJ, 499, 914
\bibitem[]{} Uitenbroek, H. 1998, ApJ, 498, 427
\end{thebibliography}
\end{document}